\documentclass[%
reprint,
amsmath,amssymb,
aps,
]{revtex4-2}
\usepackage{xcolor}
\usepackage{graphicx}% Include figure files
\usepackage{dcolumn}% Align table columns on decimal point
\graphicspath{{./figures/}}
\usepackage{enumerate}
\usepackage{subcaption}
\usepackage{bm}% bold math
\begin{document}
	
\preprint{APS/123-QED}
	
\title{Phase separation of active Brownian particles on curved surfaces} % Force line breaks with \\
	
\author{Priyanka Iyer, Roland G. Winkler, Dmitry A. Fedosov, and Gerhard Gompper}
\altaffiliation{Theoretical Physics of Living Matter, Institute of Biological Information Processing and Institute for Advanced Simulation, 
Forschungszentrum J\"ulich, 52425 J\"ulich, Germany \\ 
Email: p.iyer@fz-juelich.de, r.winkler@fz-juelich.de, d.fedosov@fz-juelich.de, g.gompper@fz-juelich.de} 
	
%\author{D. A. Fedosov}%
%\affiliation{%
%	Authors' institution and/or address\\
%	This line break forced with \textbackslash\textbackslash
%}%
	
	
%\affiliation{
%	Second institution and/or address\\
%	This line break forced% with \\
%}%
%\affiliation{
%	Third institution, the second for Charlie Author
%\affiliation{%
%	Authors' institution and/or address\\
%	This line break forced with \textbackslash\textbackslash
%}%
	
	
\date{\today}% It is always \today, today,
	
\begin{abstract}
The effect of curvature on an ensemble of repulsive active Brownian particles (ABPs) moving on a spherical surface is studied. 
Surface curvature strongly affects the dynamics of ABPs, as it introduces a new time scale $\tau=R/v_0$, with
curvature radius $R$ and propulsion velocity $v_0$, in addition to the rotational diffusion time $\tau_r$. This implies that 
motility-induced phase separation (MIPS) disappears for small $R$. Furthermore, it causes a narrowing of the MIPS regime 
in the phase diagram of P{\'e}clet number $\text{Pe}$ and particle area fraction $\phi$. Also, the phase-separation 
boundary at low $\phi$ attains a turning point at small $R$, allowing for the possibility of a reentrant behavior. 
These results characterize the effect of curvature on ABP dynamics and MIPS, and will help to better understand the 
preferred occupation of certain niches by bacterial colonies in porous media.
\end{abstract}
	

\maketitle
	
%\tableofcontents
Ensembles of self-propelling particles display rich dynamical behaviors, which arise from their out-of-equilibrium
nature \cite{bechinger2016review,elgeti2015review}. A prominent example is motility-induced phase separation 
(MIPS) in systems with no attractive or alignment interactions between the particles, which have been studied in detail 
for active Brownian particles (ABPs) both in simulations 
\cite{fily2012athermal, bialke2013theory, siebert2018critical, redner2013structure, Digregorio2018MIPS2D, nie2020stability} 
and experiments \cite{palacci2013living, buttinoni2013dynamical, liu2019self}.  
In more complex systems, several other factors have been found to affect and modify the onset 
of activity-induced clustering, such as shape anisotropy 
\cite{yang2010swarm,suma2014motility,moran2022particle,siebert2017phase,cugliandolo2017phase, Baer2020review, Schoenhoefer_CCB_2022}, 
hydrodynamic interactions \cite{theers2018clustering, worlitzer2021motility, matas2014hydrodynamic, zottl2014hydrodynamics}, 
deformability of the confinement \cite{vutukuri2020actves, takatori2020actves}, 
and dimensionality \cite{stenhammar2014phase, wysocki2014cooperative}. 
	
In biophysical systems,  active particles are often exposed to curved geometries and confinement.  
Examples include bacteria motion in porous media \cite{creppy2019porous}, cell migration on 
curved tissues of the gut \cite{ritsma2014intestinal},  embryonic development \cite{keller2008reconstruction},  
and actomyosin flows during cell division \cite{pimpale2020cell}.  
Theoretical studies of single active particles indicate that for a tangential propulsion direction, their dynamics depends 
on the surface curvature \cite{apaza2017brownian, castro2018active}, while for an unconstrained propulsion direction, 
particles predominantly accumulate in regions of higher curvature \cite{fily2015dynamics, fily2016active}.  
Steric interactions among active elongated particles, which favor polar or nematic alignment, generate complex 
flow patters on spherical geometries, such as circulating band states \cite{sknepnek2015active, li2015collective, hsu2022polar}.  
Further studies have shown topology-dependent collective dynamics of self-propelled rods \cite{janssen2017aging}, 
as well as segregation dynamics in binary mixtures of active and passive particles on spherical surfaces \cite{ai2020binary}. 
	
In this work, we study the dynamics and clustering behavior of repulsive ABPs, with a freely diffusing propulsion direction, 
but constrained to move on a curved  surface in two/three spatial dimensions (2D/3D).   
We show that the confinement radius $R$ introduces a new length scale that changes qualitatively the dynamics of ABPs on curved surfaces, such that their ballistic motion for large P{\'e}clet numbers is suppressed at distances smaller than $R$. The P{\'e}clet number is defined as $\text{Pe} = v_0\tau_r/\sigma$, where $v_0$ is the propulsion velocity, $\tau_r = D_r^{-1}$ is the rotational diffusion time with a rotational diffusion $D_r$, and $\sigma$ is the ABP diameter. The diagram of phase separation on a sphere is constructed, and shows that curvature drastically changes 
the phase boundaries, and completely suppresses MIPS at small $R$. A simple model that considers the effective persistence 
length of particle motion on the sphere is then used to rationalize the observed effects of sphere curvature on MIPS. Furthermore, we study 
MIPS in a paradigmatic example of porous media represented by two connected spheres with unequal radii.

%%%%%%%%%%%%%%%%%%%%%%%%%%%%%%%%%%%%%%%%%%%%%%%%%%%%%%%%%%%%%%%%%%%%%%%%	
\begin{figure}
\centering
\includegraphics[scale=0.98]{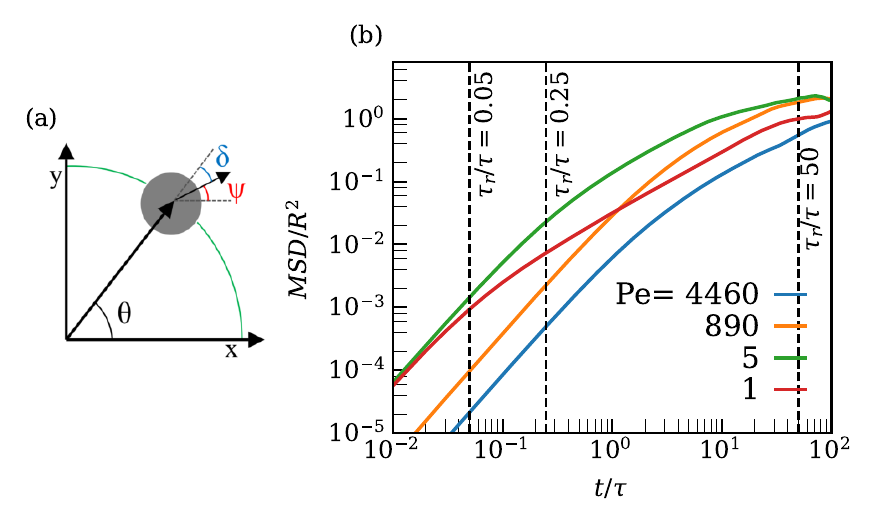}
\caption{(a) Schematic diagram of an ABP confined to a ring. 
	     (b) MSD of a particle moving on a sphere for different $\text{Pe}$ (or $\tau_r$), with $v_0$ fixed. 
	     For $\tau_r/\tau<1$, the ballistic-to-diffusive transition occurs at time $\tau_r/\tau$, whereas for 
	     $\tau_r/\tau>1$, the time scale $\tau$ determines the ballistic-to-diffusive transition irrespective of $\tau_r$.}
	\label{fig:sing_part}
\end{figure}
%%%%%%%%%%%%%%%%%%%%%%%%%%%%%%%%%%%%%%%%%%%%%%%%%%%%%%%%%%%%%%%%%%%%%%%%
	
We consider first a model of an ABP as a disc of diameter $\sigma$ in 2D, confined to a ring of radius $R$. 
While the ABP motion is restricted to one dimension, the propulsion vector ${\bf e}$ is free to rotate in 2D, 
see Fig.~\ref{fig:sing_part}(a). We neglect the effects of translational noise and focus on rotational noise. 
Then, the equations of motion for the position ${\bf r}=(R \cos\theta, R \sin\theta)$ and propulsion direction
${\bf e}=(\cos\psi, \sin\psi)$ of the ABP are 
\begin{equation}
\begin{aligned}
     \dot{\theta} = -\frac{v_0}{R} \sin(\theta-\psi), \quad \dot{\psi} = \sqrt{2D_r}\Gamma_\psi, 
\end{aligned}
\label{eq:eom}
\end{equation}
where $\Gamma_\psi$ is a Gaussian and Markovian random process with zero mean 
and $\langle \Gamma_\psi(t)\Gamma_\psi(t')\rangle = \delta(t-t')$. In the limit of small misalignment angles 
$\delta \equiv \theta-\psi$ with $|\delta| \ll 1$, Eq.~(\ref{eq:eom}) can be linearized and analytically 
solved \cite{sup_mater}, yielding the angular mean-squared displacement (MSD)  
\begin{equation}
\langle (\theta(t)-\theta(0))^2\rangle = 2D_r t - \frac{2D_r R}{v_0} (1- e^{-v_0t/R}) .
\label{eq:msd}
\end{equation}
Thus, irrespective of $\tau_r$, the crossover between the diffusive and ballistic 
regime is determined by the new time scale $\tau=R/v_0$. The shift of the onset of 
the diffusive behavior to earlier times for smaller radii $R$ originates from a fast alignment of particle orientation 
with the surface normal $\bf{r}$, as the ABP moves along the surface, upon which the translational motion of the 
particle nearly stops. Note that the MSD from Eq.~\eqref{eq:eom} agrees well with simulation results for a single ABP
on a ring \cite{sup_mater}.  

The discussion above suggests that the misalignment angle $\delta$ plays an essential role. 
The corresponding Fokker-Planck equation \cite{sup_mater} yields the stationary-state distribution
$P(\delta)\sim \exp[(\tau_r/\tau)\cos(\delta)]$, from which we obtain $\langle \delta^2 \rangle =\tau/\tau_r = R/(\sigma\text{Pe}$) 
for $\tau/\tau_r \ll 1$. For the tangential velocity $v=v_0 |\sin(\delta)|$, this implies that
$\langle v\rangle \simeq v_0 \sqrt{\tau/\tau_r}$. Thus, the ABP velocity slows down with increasing curvature 
(or decreasing $\tau$), which is also reflected in Eq.~\eqref{eq:msd} where the MSD in the ballistic 
regime is given by $R^2\langle (\theta(t)-\theta(0))^2\rangle = \langle v\rangle^2 t^2$.

The behavior of ABPs and the corresponding MSD in 3D are more complex, as both time scales $\tau$ and
$\tau_r$ become important due to additional angular degrees of freedom.
Figure~\ref{fig:sing_part}(b) shows that the ratio $\tau_r/\tau= (\sigma \text{Pe})/R$ determines 
particle dynamics. For $\tau_r/\tau\ll1$, the ABP 
does not 'see' the effect of curvature and exhibits diffusive motion for times larger than $\tau_r$. However, when $\tau_r/\tau\gg1$, 
the particle moves ballistically only up to time $\tau$.  For $t>\tau$, diffusive motion due to sphere curvature sets in, and can be described as a stop-and-go motion due to fast alignment of $\bf{e}$ along $\bf{r}$. Figure~\ref{fig:sing_part}(b) also 
demonstrates the reduction of effective particle velocity with increasing $\text{Pe}$ as the magnitude of the MSD in the ballistic regime drops. 

%%%%%%%%%%%%%%%%%%%%%%%%%%%%%%%%%%%%%%%%%%%%%%%%%%%%%%%%%%%%%%%%%%%%%%%%
\begin{figure}
\centering
	\includegraphics[scale=0.98]{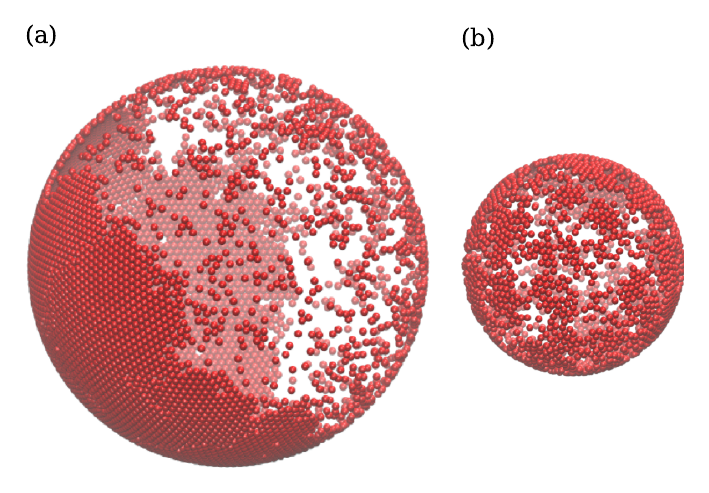}
	\caption{Simulation snapshots for different sphere radii at $\text{Pe}=890$ and $\phi=0.5$. (a) $R/\sigma=26.8$, $N=5760$, and  
           (b) $R/\sigma=16.1$, $N=2074$. A planar system with the same number of particles as in (b) exhibits phase separation \cite{sup_mater}, indicating that the absence of MIPS for $R/\sigma=16.1$ is not due to finite-size effects. See also movies S1 and S2.}
	\label{fig:sims}
\end{figure}
%%%%%%%%%%%%%%%%%%%%%%%%%%%%%%%%%%%%%%%%%%%%%%%%%%%%%%%%%%%%%%%%%%%%%%%%
	
\begin{figure*}
	\includegraphics[scale=0.93]{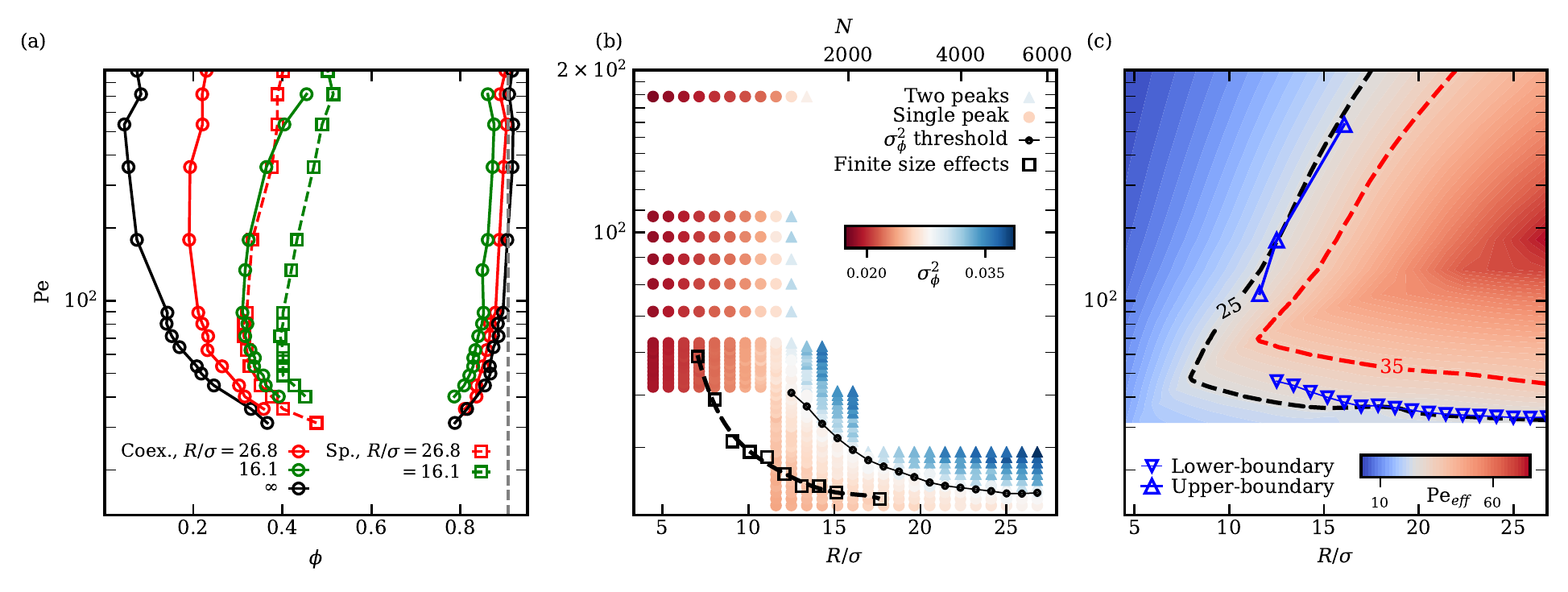}
	\caption{(a) $\text{Pe}$-$\phi$ phase diagrams of motility-induced phase separation (MIPS) for three $R$ values. 
		Coexisting densities from the local density distributions (circles) and abrupt pressure drops (squares) are
		employed to identify the transition. The simulations for determining the coexisting densities are performed 
                for an average area fraction of $\phi=0.5$. 
		(b) Critical P{\'e}clet number $\text{Pe}_c(R)$ at which MIPS is first observed for increasing $\text{Pe}$ at fixed $\phi=0.5$. 
		The color-bar shows the variance of the local density distribution. Symbols mark the identification of 
		no-MIPS (uni-modal $P(\phi_{loc})$) (circles) and MIPS (bi-modal $P(\phi_{loc})$) (triangles). 
		The black line with bullets is obtained from the threshold $\sigma^2_\phi=0.0305$ of variance of the local density, and
		it follows well the boundary where the two peaks in the local density distribution merge. 
        (c) Heat map of the effective P{\'e}clet number $\text{Pe}_{eff}$ for a single ABP at the sphere surface as a function 
        of $\text{Pe}$ and $R$. The black and red dashed lines represent $\text{Pe}_{eff} = 25$ and $\text{Pe}_{eff} = 35$, and 
		match well the lower and upper MIPS boundaries from the simulations.
		}
	\label{fig:MIPS_radius}
\end{figure*}

We consider next an ensemble of $N$ ABPs on a sphere with area packing fraction $\phi$, to study how MIPS is affected 
by the sphere curvature $1/R$. In the simulations \cite{sup_mater}, $\text{Pe}$ is changed by varying $\tau_r$ while $v_0$ is kept fixed. 
Figure~\ref{fig:sims} shows simulation snapshots for two different curvatures at $\text{Pe}=890$, and demonstrates the absence of 
MIPS for the small radius $R/\sigma = 16.1$. The full phase diagram for different $R$ values is presented 
in Fig.~\ref{fig:MIPS_radius}(a).  Here, the {\em binodal} is constructed by measuring co-existing densities in the phase 
separated state, whereas the {\em spinodal} is obtained by computing the particle pressure \cite{sup_mater, solon2018generalized}. 
A sudden drop/change in pressure marks the transition from the homogeneous to the phase-separated state
\cite{wittkowski2014scalar,winkler2015virial,levis2017active,solon2018generalized}. Two main effects of curvature 
can be seen in Fig.~\ref{fig:MIPS_radius}(a) for decreasing $R$: (i) the lower part of binodals and spinodals 
shifts to larger $\text{Pe}$, and (ii) the two-phase region becomes narrower and the slope of the left spinodal/binodal changes sign for 
large $\text{Pe}$.

Figure~\ref{fig:MIPS_radius}(b) shows the variation of the critical P{\'e}clet number $\text{Pe}_c(R)$, where MIPS is first observed 
(binodal) with increasing $\text{Pe}$ for a fixed initial density of $\phi=0.5$. Since the value of $R$ sets the total 
number $N$ of particles for a given $\phi$, the system may inherently suffer from 
finite-size effects. To account for the possible effects of finite $N$, planar simulations with the 
same number of ABPs are performed, see  Fig.~\ref{fig:MIPS_radius}(b). $\text{Pe}_c$ increases with 
decreasing $N$ in both cases, see Fig.~\ref{fig:MIPS_radius}(b). However, this effect is much less pronounced
for the planar systems, demonstrating that finite-size effects are sub-dominant.
Furthermore, the width of the MIPS region becomes narrower with increasing $\text{Pe}$ [see Fig.~\ref{fig:MIPS_radius}(a)], 
which explains the sudden disappearance of MIPS at $R/\sigma \approx 12.5$ and $\phi=0.5$ in Fig.~\ref{fig:MIPS_radius}(b) 
(i.e., the spinodal has a turning point before $\phi=0.5$ is reached for $R/\sigma \lesssim 12.5$). This clearly implies that the loss of MIPS at $R/\sigma \approx 12.5$ is not due to 
finite-size effects.

%%%%%%%%%%%%%%%%%%%%%%%%%%%%%%%%%%%%%%%%%%%%%%%%%%%%%%%%%%%%%%%%%%%%%%%%
%%%%%%%%%%%%%%%%%%%%%%%%%%%%%%%%%%%%%%%%%%%%%%%%%%%%%%%%%%%%%%%%%%%%%%%%
	
MIPS occurs as a result of slowing down of ABPs due to crowding, which promotes a further reduction in velocity and clustering 
through a positive feedback mechanism \cite{buttinoni2013dynamical,Cates_MIPS_2015}. A requirement for MIPS is that the 
life time of small clusters is larger than the persistent travel time of ABPs \cite{bruss2018phase,matas2014hydrodynamic}. 
This means that the directed self-propelled motion should dominate over diffusive motion on the length scale of particle
diameter $\sigma$, i.e. $\sigma/\tau_r \ll v_0$ or $\text{Pe}=v_0 \tau_r/\sigma \gg 1$. 
For ABP motion on a curved surface, this argument has to be modified as follows. First, the propulsion velocity $v_0$ has to
be replaced by a radius-dependent velocity $v(R)$. In general, $v(R)$ decreases 
with decreasing $R$, e.g., in 2D, $v(R)=v_0\sqrt{\tau/\tau_r}=v_0\sqrt{R/\text{Pe}}$ for $R/\text{Pe}<1$. 
Second, on a curved surface, the time scale $\tau$ becomes
relevant in addition to $\tau_r$. Thus, we have to distinguish the two cases $\alpha\tau > \tau_r$ and $\alpha\tau < \tau_r$ ($\alpha$ is a constant of order unity), which represent large and small radii $R$, respectively. 
In both cases, the shorter time scale $\tau_{min} = \min(\alpha\tau,\tau_r)$ determines the dynamics [see Figure~\ref{fig:sing_part}(b)]. 
As a result, we can define an effective P{\'e}clet number $\text{Pe}_{eff} = v(R)\tau_{min}/\sigma$, which has to exceed the threshold $\text{Pe}_c$ for phase separation to occur. Hence, a larger bare $\text{Pe}$ is required for a smaller $R$ to compensate for the reduced effective surface velocity.  
	
From this argument, all the trends observed in Fig.~\ref{fig:MIPS_radius} can be understood. With decreasing $R$, 
$\text{Pe}_c$ first increases, because $\tau_r$ is the relevant time scale and $v(R)$ decreases, and this increase has to be compensated 
by a larger $\text{Pe}$. Note that $\text{Pe}_{eff}$ increases with increasing $\tau_r$ and $v_0$ only as long as $\tau_r<\alpha \tau$. 
When $\tau_r=\alpha\tau$, $\text{Pe}_{eff}$ reaches a maximum as a function of $\text{Pe}$ for a fixed $R$. A further increase 
of $\text{Pe}$ only causes a decrease in the effective surface velocity $v(R)$, without any increase in $\tau_{min}=\alpha\tau$. 
This leads to a decrease in $\text{Pe}_{eff}$ and the turning of the low-$\phi$ branch of two-phase coexistence toward larger 
$\phi$ values in Fig.~\ref{fig:MIPS_radius}(a).
	
At low-to-intermediate particle densities ($\phi<0.5$) and smaller radii, with $\tau_r>\alpha\tau$, MIPS is absent for all $\text{Pe}$.  
This inversion of time scales and disappearance of MIPS occur when particle diffusion dominates over  
the minimum run length for cluster formation. In this case, an increase of $\tau_r$ cannot lead to MIPS, because the
slowing down due to translational ABP motion on a curved surface always precedes rotational diffusion. 
Furthermore, this argument indicates that for smaller $R$ values, the occurrence of MIPS requires larger area fractions $\phi$ 
because $\text{Pe}_c$ is larger, as shown in Fig.~\ref{fig:MIPS_radius}(a) where MIPS for $R/ \sigma=16.1$ occurs 
only at $\phi \gtrsim 0.35$. This curvature effect further lowers the range of $R$ where MIPS is observed.  
	
To verify that the effective P{\'e}clet number $\text{Pe}_{eff}$ indeed controls phase separation on curved surfaces, 
average surface velocity $v(R)$ for a single ABP is measured in simulations. Figure~\ref{fig:MIPS_radius}(c) shows a 
heat map of $\text{Pe}_{eff}$ for various radii, where $\alpha=6$ is selected for a good fit of the simulation 
data for MIPS. Lower and upper boundaries of the MIPS region for a fixed $\phi = 0.5$ agree well with the black dashed 
line for $\text{Pe}_{eff} = 25$. Furthermore, the heat map of $\text{Pe}_{eff}$ nicely explains the loss of MIPS at small 
$R$. Figure~\ref{fig:MIPS_radius}(c) also shows that for a larger critical $\text{Pe}_c$ (or smaller $R$), the MIPS 
regime becomes narrower as a function of $R$, consistent with the onset of MIPS at larger $\phi$. Noteworthy, 
$\text{Pe}_{eff}$ reaches a maximum and then decreases as a function of $\text{Pe}$, which explains the turning of the 
phase boundary at large $\text{Pe}$. This supports the existence of a {\em reentrant} behavior (i.e., from homogeneous 
to MIPS and back to homogeneous density) with increasing $\text{Pe}$ for a wide range of radii. Note that we have not 
observed any significant change in the right binodal/spinodal at large $\phi$ with curvature. This is due to high 
densities, at which inter-particle collisions are very frequent, so that the particle dynamics is significantly 
affected and the simple estimate based on single-particle $\text{Pe}_{eff}$ is not valid.

%%%%%%%%%%%%%%%%%%%%%%%%%%%%%%%%%%%%%%%%%%%%%%%%%%%%%%%%%%%%%%%%%%%%%%%%
\begin{figure}
	\centering
	\includegraphics[scale=0.98]{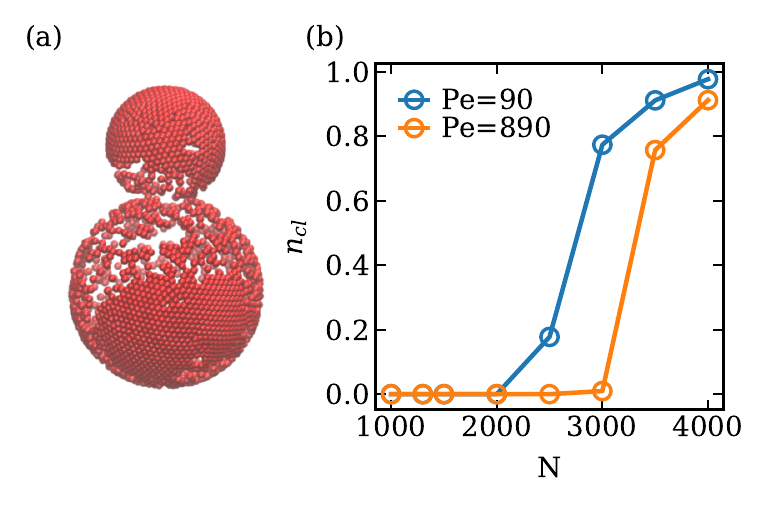}
	\caption{(a) Snapshot of a system of two interconnected pores of radii $R_1/\sigma=16.1$ and $R_2=0.6R_1$ 
		for $N=2500$ at $\text{Pe}=90$ (see also movie S3). (b) Fraction $n_{cl}$ of particles in the large sphere occupying clusters of size 
		greater than $N_1/2$ as a function of $N$. The sudden jump identifies MIPS. }
	\label{fig:two_sph_main}
\end{figure}
%%%%%%%%%%%%%%%%%%%%%%%%%%%%%%%%%%%%%%%%%%%%%%%%%%%%%%%%%%%%%%%%%%%%%%%%
	
Following the characterization of the behavior of ABPs at a surface with a fixed curvature, we make a further step toward the understanding of their behavior in porous media. We consider a paradigmatic example of two spherical pores with unequal radii, which are connected by a small passage as shown in Fig.~\ref{fig:two_sph_main}(a). For low area densities of ABPs, the solution of the Langevin equation in a convex, non-spherical confinement in 3D shows that the 
single-particle density is proportional to the squared local curvature of the boundary \cite{fily2015dynamics, fily2016active}. 
For two connected pores in 3D, this implies that the ratio of particle area fractions $\phi_1$ and $\phi_2$ within the larger and smaller spheres with radii $R_1$ and $R_2$ is given by $\phi_1/\phi_2 = (R_2/R_1)^2$. Therefore, for non-interacting ABPs at the steady state, their area fraction in the smaller pore is larger than that in the larger pore, while the average number $N_i$ of 
particles in each pore $i=1,2$ should be the same, i.e. $N_1=N_2=N/2$. As the (equilibrium) particle pressure with excluded-volume 
interactions increases faster with area fraction than for an ideal gas, the particle number $N_2$ in the smaller pore 
should be smaller than $N/2$, and correspondingly $N_1 > N/2$. Then, our results in Fig.~\ref{fig:MIPS_radius}(a) for MIPS 
in a single pore allow for predictions of the steady-state behavior of ABPs in the two pores. As $N$ is increased for a 
given $\text{Pe}$, the particles in either sphere can phase separate only when their surface density exceeds $\phi_c(R,\text{Pe})$ 
for MIPS. The density in the smaller pore, which fills first, increases linearly with $N$, and MIPS is expected at
$N=2\phi_c(R_2,\text{Pe})A_{2}/A_{\sigma}$, where $A_2$ is the area of the 
smaller sphere minus the area of the passage between two pores, and $A_{\sigma}$ is the area occupied by an ABP. The number of particles in the smaller pore cannot exceed 
$N_{2,max}=\phi_{cp}A_{2}/A_{\sigma}$, where $\phi_{cp}$ is the close-packing density. Therefore, the number 
of particles in the larger pore at the steady state is expected to be $N_1(N)=N-\text{min}(N/2,N_{2,max})$. As a result, for a 
given $\text{Pe}$, phase separation in the larger sphere is first expected to occur near the value of $N$ that satisfies the 
equality $N_1(N)=\phi_c(R_1,\text{Pe})A_{1}/A_{\sigma}$ ($A_1$ is the area of the larger 
sphere minus the passage area).

To test our predictions, we consider two connected pores with $R_1/ \sigma =16.1$ and $R_2/R_1=0.6$, where MIPS in the larger pore 
is expected at $N\simeq2900$ and $N\simeq3300$ for $\text{Pe}=90$ and $\text{Pe}=890$, respectively. 
These predictions are nicely confirmed by the simulation results in Fig.~\ref{fig:two_sph_main}(b), where the onset of phase 
separation as a function of $N$ is characterized by a sudden rise in the fraction $n_{cl}$ of particles occupying clusters 
of size greater than $N_1/2$. Note that the reduction of the effective P{\'e}clet number at large $\text{Pe}$ implies that a 
larger $N$ is required for MIPS. 

Another interesting observation is the temporal evolution of this system after starting with a uniform equilibrium distribution of ABPs 
and approaching the transition regime. For sufficiently large $N$, the larger 
pore shows a 'dynamic' MIPS state. At short times, the initial area fraction is large enough to show MIPS, 
which eventually dissolves as ABPs are lost to the smaller pore with time. For low P{\'e}clet numbers, due to 
large fluctuations in the particle number in the larger sphere, the MIPS state can both be dynamically restored and 
lost in time \cite{sup_mater}. 

In summary, if the propulsion direction of ABPs can vary diffusively (i.e., it is not aligned with the local tangent plane 
of a surface), the behavior of active particles on curved surfaces is very different in comparison to ABPs confined to a plane. 
Non-zero curvature results in a `stop-and-go' motion, such that particles slide along the surface when their orientation is 
different from the local normal, and then stop after their orientation becomes perpendicular to the surface. 
This behavior governs motility-induced phase separation on curved surfaces, e.g., the MIPS region rapidly shrinks with 
increasing curvature and eventually disappears. Furthermore, curved surfaces lead to a possible reentrant behavior, where 
MIPS for a fixed surface density of ABPs first appears with increasing $\text{Pe}$, and then can disappear. The single-pore 
results also allow us to predict the dynamics of active particles in connected pores with distinct curvatures as in porous 
media. These results will help to better understand the preferred occupation of certain geometries and niches by bacterial 
colonies \cite{Chang_BFF_2015}.

%\bibliography{manuscriptNotes}% Produces the bibliography via BibTeX.


\end{document}